\documentclass[multphys,vecphys]{svmult}

\usepackage{makeidx}         
\usepackage{graphicx}        
\usepackage{multicol}        
\usepackage[bottom]{footmisc}
\fontsize{12pt}{0} 
\selectfont

\makeindex             

\begin{document}

\title*{Transition radiation in turbulent astrophysical medium. Application to solar radio bursts.}
\author{Gregory D. Fleishman\inst{1,2}, Dale E. Gary\inst{1}, Gelu M. Nita\inst{1}} 
\institute{$^1$New Jersey Institute of Technology, Newark, NJ 09102 \\
$^2$Ioffe Physico-Technical
Institute, St. Petersburg, 194021 Russia \\
}
%

%
\maketitle

\abstract{Modern observations and models of various astrophysical
objects suggest that many of their physical parameters fluctuate
substantially at different spatial scales. The rich variety of the
emission processes, including Transition Radiation but not limited
to it, arising in such turbulent media constitutes the scope of
Stochastic Theory of Radiation. We review general approaches applied
in the stochastic theory of radiation and specific methods used to
calculate the transition radiation produced by fast particles in the
magnetized randomly inhomogeneous plasma. The importance of the
theory of transition radiation for astrophysics is illustrated by
one example of its detailed application to a solar radio burst,
including specially designed algorithms of the spectral forward
fitting.}

\def\gsim{\ \raise 3pt \hbox{$>$} \kern -8.5pt \raise -2pt \hbox{$\sim$}\ }
\def\lsim{\ \raise 3pt \hbox{$<$} \kern -8.5pt \raise -2pt \hbox{$\sim$}\ }

\section{Introduction}

 The phenomenon of transition radiation was discovered
theoretically by two Nobel Prize winning (2003 and 1958
respectively) physicists \cite{Gin_Fr}. Ginzburg and Frank (1946)
considered a simplest case when a charged particle passed through a
boundary between two dielectrically different media and so generated
waves due to a variation of the dielectric constant at the boundary.
Remarkably, no acceleration of the particle is necessary to produce
the emission due to transition through the boundary.

It is easy to understand that a similar effect of electromagnetic
emission will take place if a medium is uniformly filled by
turbulence that produces fluctuations of the dielectric constant
throughout the whole volume rather than at an isolated boundary.
Many astrophysical sources, especially those under strong energy
release, are believed to be filled by turbulent, randomly
inhomogeneous plasma and fast, nonthermal particles. In this
situation, an efficient contribution of the transition radiation to
the overall electromagnetic emission should be produced. Therefore,
distinguishing this contribution from competing mechanisms is
important. Below we describe the fundamentals of the transition
radiation produced in a magnetized turbulent plasma, and demonstrate
its high potential for astrophysical applications.

\section{ Fundamentals of Stochastic Theory of Radiation. }

Stochastic theory of radiation now represents a broad field of
physics with many applications. In astrophysical sources the
nonthermal radiation arises as charged fast particles move through a
turbulent plasma with random fluctuations of plasma density,
electric, and magnetic fields. An immediate consequence of the
random magnetic and electric fields at the source is that the
trajectory of a charged particle is a random function of time. This
means that calculating the emission requires some appropriate
averaging of the relevant equations and parameters over the possible
particle paths.

The presence of the density inhomogeneities acts differently.
Indeed, these inhomogeneities have little effect on the fast
particle trajectories; rather they give rise to fluctuations of the
plasma dielectric tensor. These fluctuations allows for the plasma
current stimulated by the fast particle field to emit powerful
transition radiation. Since the plasma has a resonance around the
plasma frequency, the intensity of the transition radiation is
extremely large around this plasma frequency. The corresponding peak
in the transition radiation spectrum is referred to as resonant
transition radiation (RTR) and is of exceptional importance for
astrophysical applications. Indeed, the microturbulence accompanied
by the density fluctuations is likely to exist in various cosmic
objects from geospace to distant cosmological sources of gamma ray
bursts. In many instances, however, the number density of the plasma
in the astrophysical object and the corresponding plasma frequency
is so low that the radiation cannot be observed at the Earth because
of ionosphere opacity and absorption of this radiation in the
interstellar medium.

In some cases, nevertheless, the plasma frequency of a source is
large enough for the corresponding radiation to be observable. For
example, this is true for radio bursts produced in the solar corona.
In our previous publications \cite{RTR,RTR_letter} we demonstrated
that about 10$\%$ of all microwave solar continuum bursts are
accompanied by decimetric resonance transition radiation (RTR) and
presented ample evidence in favor of transition radiation for one of
the events, 06 April 2001, including detailed study of spatially
resolved observations.

Specifically, Nita et al. (2005) summarized and checked against
observations the following main properties of RTR, expected in the
case of solar bursts. The emission (1) originates in a dense plasma,
$f_{pe} \gg f_{Be}$, where $f_{pe}$ and $f_{Be}$ are the electron
plasma- and gyro-frequencies; (2) has a relatively low peak
frequency in the decimetric range, and so appears as a low-frequency
component relative to the associated gyrosynchrotron spectrum; (3)
is co-spatial with or adjacent to the associated gyrosynchrotron
source; (4) varies with a time scale comparable to the accompanying
gyrosynchrotron emission (assuming a constant or slowly varying
level of the necessary microturbulence); (5) is typically strongly
polarized in the ordinary mode (o-mode), since the extraordinary
mode (x-mode) is evanescent, as for any radiation produced at the
plasma frequency in a magnetized plasma; (6) is produced by the
lower-energy end of the same nonthermal electron distribution that
produces the gyrosynchrotron emission, with intensity proportional
to the instantaneous total number of the low-energy electrons in the
source; (7) has a high-frequency spectral slope that does not
correlate with the spectral index of fast electrons (in contrast to
gyrosynchrotron radiation, which does). Here we analyze this event
in even more detail and apply the observations to the flare plasma
diagnostics.

\section{  Two-component radio burst 06 April 2001. }

Fig. 1 gives an overview of the event under study. It displays the
dynamic spectrum of the radio burst 06 April 2001 recorded by Owens
Valley Solar Array (OVSA) in the frequency range 1-18 GHz (Nita et
al. 2005) in total intensity and polarization, as well as images of
the dm and cm sources superimposed on a Transition Region and
Coronal Explorer (TRACE) 171 \AA\ image. This figure shows that the
event indeed consists of two distinct spectral components, whose
sources coincide spatially with each other and with a dense loop
visible in the TRACE UV image. The dm component is highly right-hand
circularly (RCP) polarized.

\begin{figure} [htbp]
\includegraphics[height=2.03in]{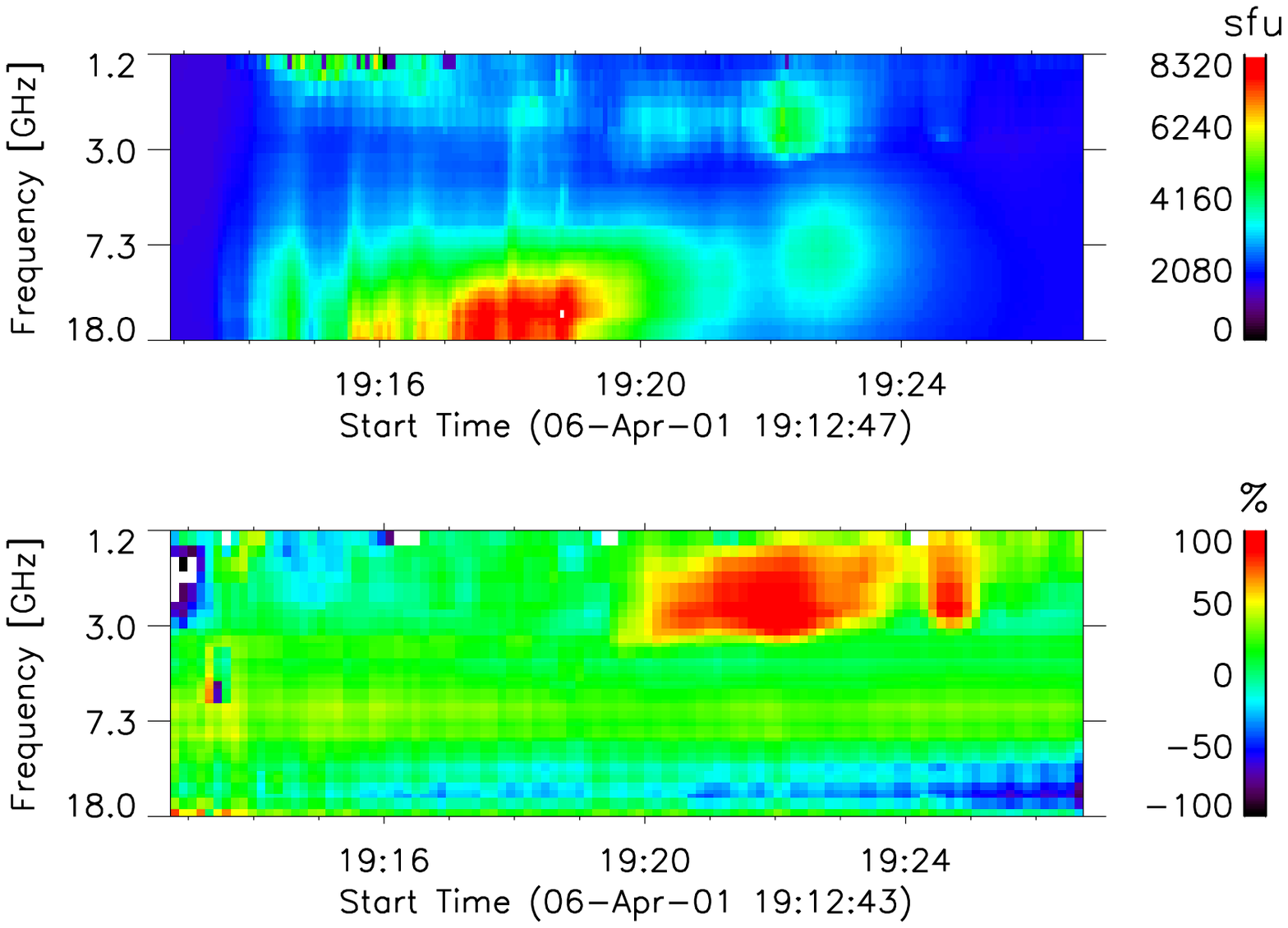}
\includegraphics[height=2.03in]{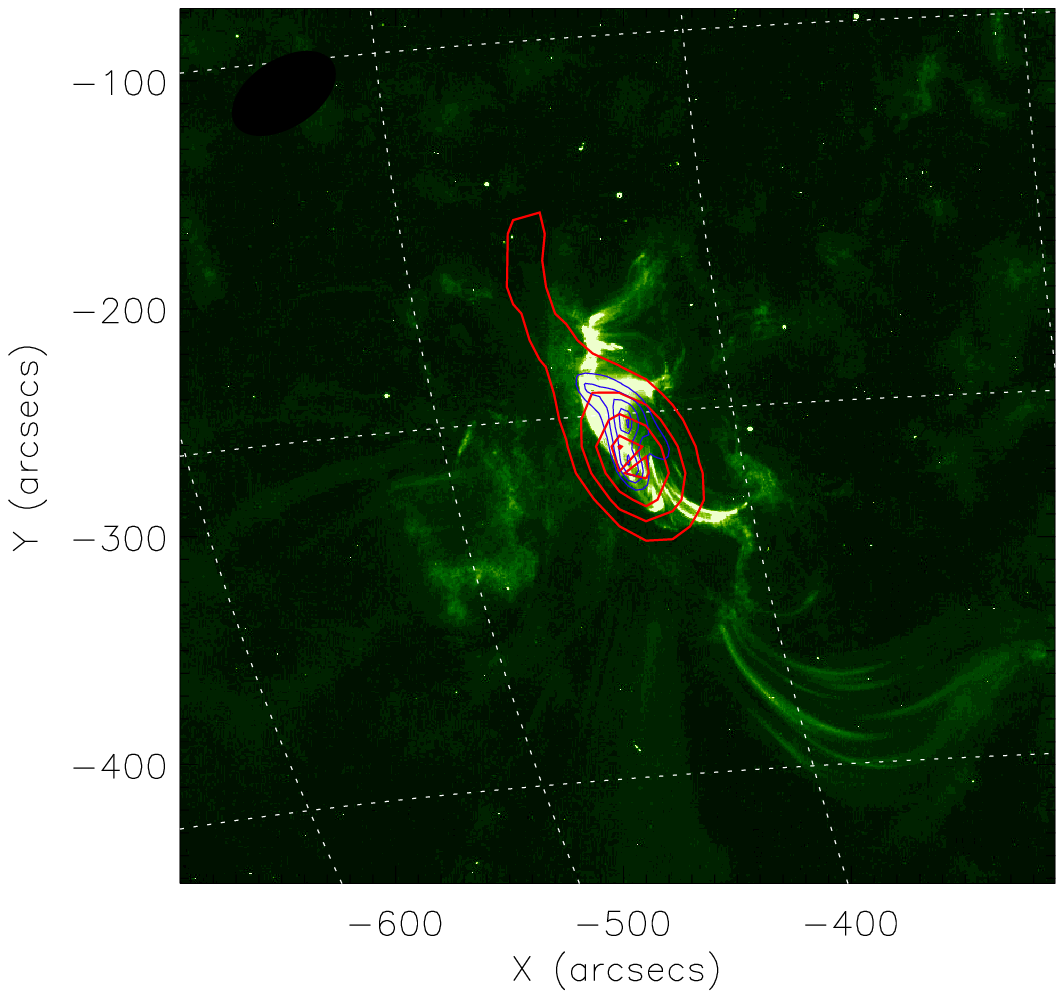}
\caption{\small Overview of the event. Left: dynamic spectra of the
total power (top) and circular polarization (bottom). The period of
RTR is the highly polarized (red) emission in the lower panel.
Right: OVSA maps of radio emission at 2 GHz (red) and 7.4 GHz (blue)
superimposed on the TRACE 171 \AA\ image.} \label{fig1}
\end{figure}

Fig. 2 displays contour levels of the dm emission at 95, 97 and
99$\%$ of the peak intensity at eight dm OVSA frequencies in RCP
(thick lines) and LCP (thin lines) superimposed on the photospheric
magnetogram (left) and on the SXR SOHO image of a dense hot loop
(right).  A number of things can be noted in this figure. The
stronger, RCP, component of the dm emission originates from the
region of the negative magnetic polarity throughout all the dm
frequency channels; therefore, it is O-mode radiation. All the RCP
contours coincide spatially with the brightest part of the hot dense
loop visible in SXR, i.e., this emission goes from a region with
relatively large plasma density, although exact position of the peak
brightness changes with frequency. The trend of this change is such
that the lower frequency radiation tends to originate at the
loop-top, while higher frequency radiation tends to originate from
the loop legs. The behavior of the weaker LCP component bears both
similarities and differences with the RCP component. First, the LCP
emission at 1.2-2 GHz comes from the region of positive magnetic
polarity, so it is also O-polarized radiation. Second, the lower
frequency LCP sources are displaced relative to the brightest part
of the SXR loop. And finally, the higher frequency LCP sources (at
2.4-3.2 GHz) are located in a region with negative magnetic
polarity; therefore, it is X-mode polarized radiation, unlike other
dm radiation. These spatial relationships are in excellent agreement
with expectations derived from the spectral behavior of this radio
burst: indeed, since the LCP contribution of the RTR component is
very weak, the LCP emission at 2.4-3.2 GHz is interpreted as a
continuation of the X-polarized microwave gyrosynchrotron radiation,
rather than RTR contribution.

\begin{figure} [htbp]
\includegraphics[height=3.03in]{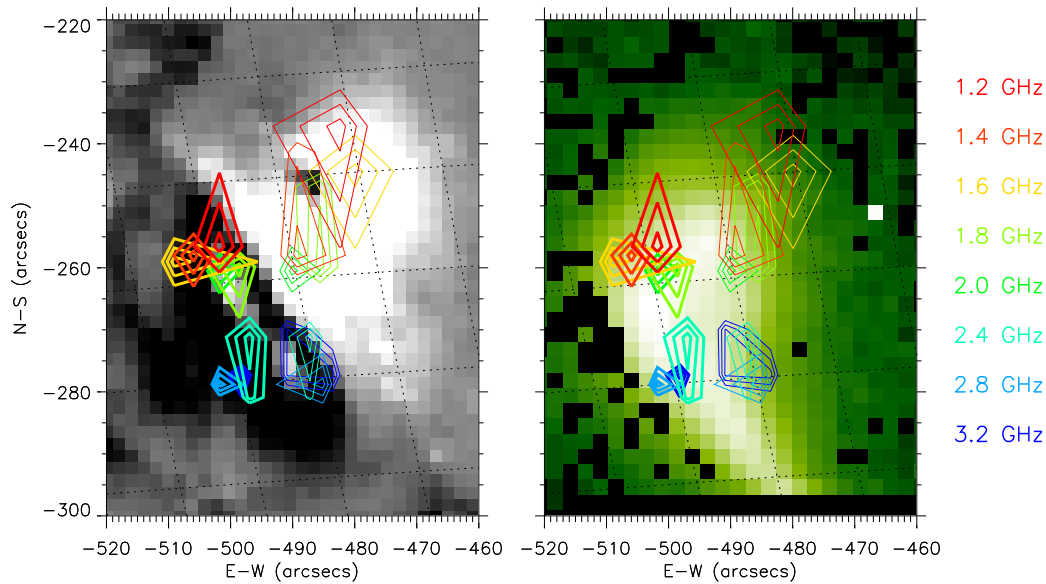}
\caption{\small OVSA maps of the RCP (thick lines) and the LCP (thin
lines) dm emission at 8 distinct frequencies superimposed on the
magnetic field distribution (left) and SXR SOHO map (right).}
\label{fig2}
\end{figure}

In addition to previously established properties of the dm continuum
component (Nita et al. 2005) and spatial relationships discussed
above, we consider here the characteristic decay time constants at
both dm and cm spectral components. Specifically, we looked into the
decay phases of the light curves at all frequencies and determined
the range of time when the decay profile can be approximated by an
exponential function. Then, the characteristic decay constant at
each light curve was considered as a characteristic decay time at
this frequency. Fig. 3 displays these time decay constants vs
frequency for the dm and cm components separately. The dependences
are remarkably different for these two emission components, although
both of them can be easily understood if the decay constants are
specified by the fast electron life times against the Coulomb
collisions at the source. Indeed, for the gyrosynchrotron emission,
higher emission frequency means greater mean energy of the emitting
electrons. These higher energy electrons have a longer life time
against the Coulomb collisions, which explains the observed increase
of the decay constant with frequency. For the RTR contribution,
however, higher emission frequency corresponds to a higher plasma
frequency in the corresponding source level, therefore, denser
plasma and, consequently, shorter life time of the fast electrons,
which is really observed. Thus, the more detailed study performed
here confirms further the interpretation of the dm continuum
component as an RTR contribution.

\begin{figure} [htbp]
\includegraphics[height=3.03in]{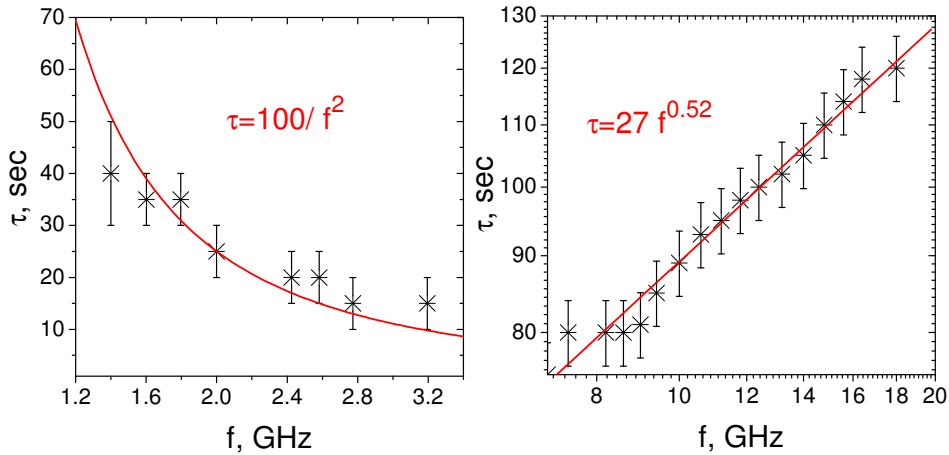}
\caption{\small The decay constant of the radio emission vs
frequency for the dm component (left) and cm component (right).}
\label{fig3}
\end{figure}

\section{Flare plasma diagnostics based on the transition radiation.}

Now, as we have a solid interpretation of the dm continuum component
as produced by RTR by fast electrons moving in a dense plasma, we
can make a next step and apply the observations for the flaring
plasma diagnostics. We assume that the inhomogeneity of the flare
volume can be described by a Gaussian distribution of the source
volume over the plasma frequency with some mean plasma frequency
(mean plasma density) and dispersion (provided by scatter of the
plasma density through the source), $F(f_{pe})=A \exp(-(f_{pe} -
f_0)^2/\Delta f^2)$. Depending on how other relevant parameters
change with the plasma density at the source, the RTR intensity can
be parameterized as proportional to $f^a \exp(-(f - f_0)^2/\Delta
f^2)$ with different values of the parameter $a$. For clarity, we
use two distinct values $a$=0 (model 1 -- no dependence on frequency
besides the Gaussian factor) and $a$=2 (model 2 -- rather strong
frequency dependence).

Then, we fit the sequence of the observed dm spectra with these two
model functions to obtain a sequence of the fitting parameters,
which are the peak flux, the mean plasma density $f_0$, and the
dispersion $\Delta f$.  Fig. 4 displays the sequence of the recorded
spectra and corresponding fits. It is clear that fits with both
values of $a$ are comparably good and even indistinguishable for
most of the instances.

\begin{figure} [htbp]
\includegraphics[height=6.03in]{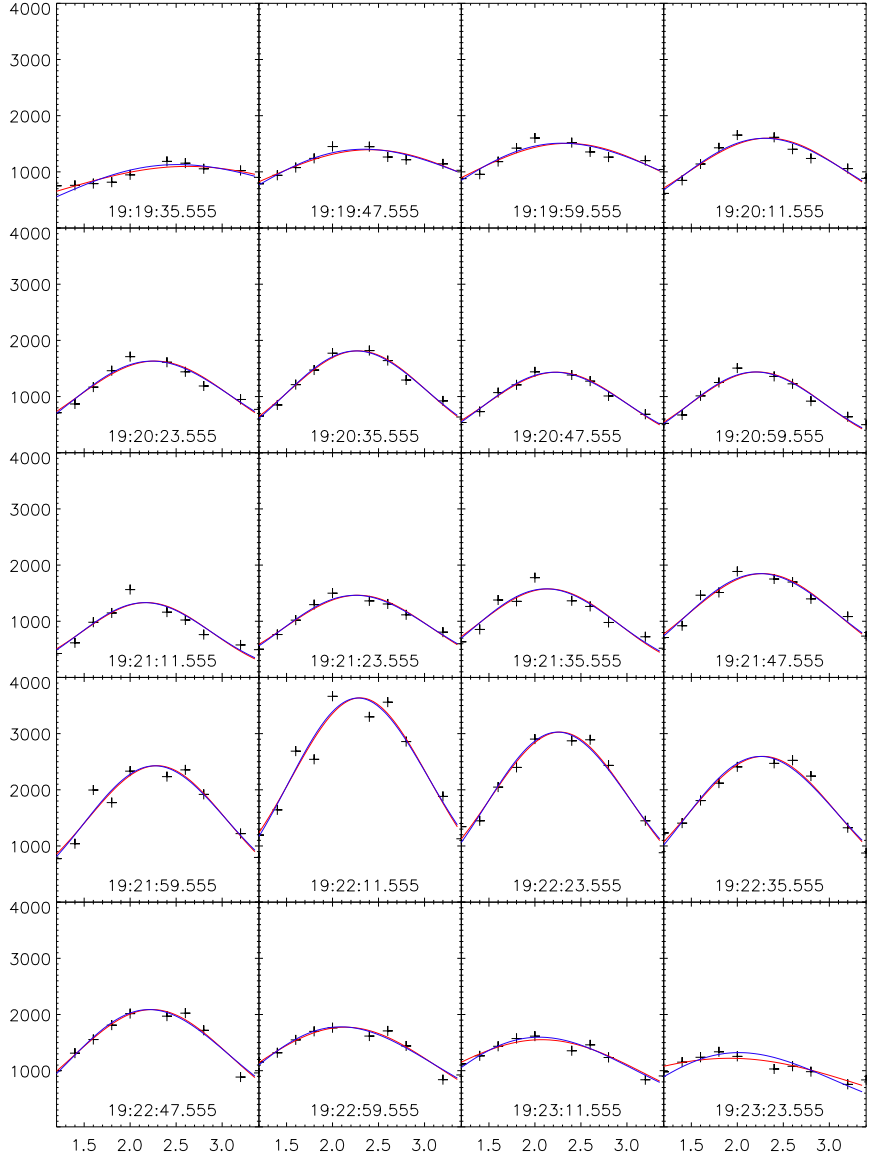}
\caption{\small Sequence of the dm spectra with two fits: S(f) ~ fa
exp(-(f - f0)2/?f2). Model 1 (a=0, red) and model 2 (a=2, blue). }
\label{fig4}
\end{figure}

Fig. 5 gives the sequence of the fitting parameters together with
the chi-square plot. Again, we cannot select between the two models
based on the chi-square criterion, since they are very similar to
both considered models. Both models give the same peak flux values
and similar bandwidth of the distributions. However, the central
frequency of the distribution behaves differently in the two models:
it is almost constant with 10$\%$ variations in model 1, while it
varies substantially in model 2. Thus, using the requirement of
reasonable smoothness of the derived physical parameters we conclude
the model 1 (which assumes that all other parameters besides the
plasma density are constant through the source) is preferable.

\begin{figure} [htbp]
\includegraphics[height=5.03in]{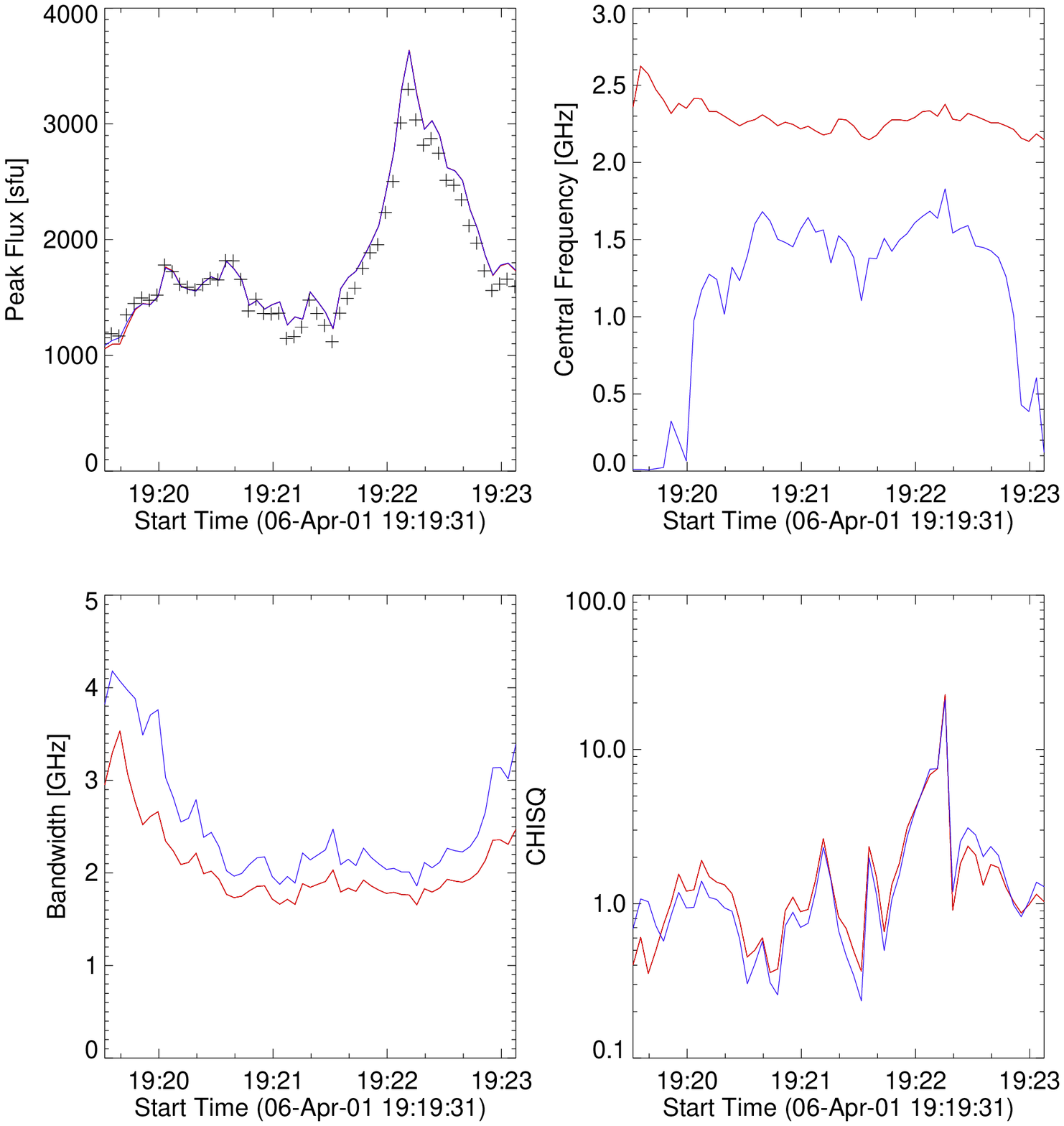}
\caption{\small Evolution of the fitting parameters and chi-square
measure for model 1 (red) and model 2 (blue). } \label{fig5}
\end{figure}

\section{Discussion}

Nita et al. (2005) proved that the dm continuum component of this
solar radio burst is produced by RTR and derived the level of the
microturbulence in the plasma to be $\left<\Delta n^2\right>/n^2 =
10^{-5}$. This finding is potentially very important for other
cosmic objects. Indeed, the obtained microturbulence level is not
particularly strong and much stronger turbulence is expected in many
cases, especially, when there is a strong release of the energy at
the source. Sometimes, such energy release gives rise to a
relativistic expansion of the source, so the emission spectrum is
Doppler-boosted and RTR produced at the local plasma frequency can
be observed at the Earth even from relatively tenuous sources with
low plasma frequency.

In this study we present more evidence in favor of RTR generation at
the dm continuum solar bursts and use this emission component to
derive additional plasma parameters. In particular, we determine the
mean plasma frequency and its dispersion at the source in the course
of time. Interestingly, these two parameters do not change much
during the time of the dm burst. We note that these parameters are
obtained from the total power spectra recorded without spatial
resolution. Fig. 2 demonstrates that the radio sources at various
frequencies do not coincide exactly. Therefore, in cases where a
sequence of spatially resolved spectra are available we would be
able to study the structure of the flaring plasma density in much
greater detail as well as the distribution of the microturbulence
over the source.

Generally speaking, the RTR contribution is also informative about
the fast electrons producing it. In the example presented in Fig. 3
we show the emission decay constants, which can be associated with
the fast electron life times. In our case we used exponential
fragments of the light curves at the late decay phase of the
emission, since no exponential phase was found in the early decay
phase. We found the life time to be within 10-40 sec, which
corresponds to the electrons of 300 keV or larger in the case of
dense flare plasma available in this event. On the other hand, we
can expect that most of the RTR emission (around the peak of the
burst) is produced by the electrons with E=100-200 keV (Nita et al.
2005). This apparent contradiction can be easily resolved if we
recall that the lower energy electrons have the life time of only a
few seconds in the given dense plasma, so they die even before the
light curves reach the exponential decay stage, and we observe the
RTR contribution from preferentially higher energy electrons late in
the event.

\section*{Conclusions.}

RTR represents a distinct emission mechanism, which is observable in
many solar burst and probably relevant to many other cosmic sources.
Application of the available theory to observations allows for
advanced plasma diagnostics, including study of the plasma density
and the turbulence distributions, and fast particle kinetics.  The
theory of RTR is currently developed for usual plasmas, but is not
available for the case of relativistic plasmas. Given that the
turbulence level in the relativistic plasmas is expected to be very
high, the extension of the RTR theory to the relativistic case is
highly desirable.

\textbf{Acknowledgments.} GDF is grateful to organizers of the
Baikal School for Fundamental Physics for invitation to deliver a
lecture on transition radiation. This work was supported in part by
NSF grant ATM-0707319 to New Jersey Institute of Technology, and by
the Russian Foundation for Basic Research, grants No. 06-02-16295,
06-02-16859, 06-02-39029.

\end{document}